\documentclass{article}

\usepackage{arxiv}

\usepackage[utf8]{inputenc} 
\usepackage[T1]{fontenc}    
\usepackage{hyperref}       
\usepackage{url}            
\usepackage{booktabs}       
\usepackage{amsfonts}       
\usepackage{nicefrac}       
\usepackage{microtype}      
\usepackage{graphicx}
\usepackage{doi}

\newcommand{\genmotion}{{\sc GenMotion}}

\title{GenMotion: Data-driven Motion Generators for Real-time Animation Synthesis}

\author{ Yizhou Zhao \\
	Department of Statistics\\
	University of California, Los Angeles\\
	\texttt{yizhouzhao@g.ucla.edu} \\
	\And
	 Wensi Ai \\
    Department of Electrical and Computer Engineering\\
    University of California, Los Angeles\\
	\texttt{ va0817@g.ucla.edu} \\
	\And
	Liang Qiu \\
	 Department of Electrical and Computer Engineering\\
	University of California, Los Angeles\\
	\texttt{liangqiu@g.ucla.edu} \\
	\And
	Pan Lu \\
    Department of Computer Science\\
    University of California, Los Angeles\\
	\texttt{panbruin@g.ucla.edu } \\
		\And
	Shi Feng \\
    Department of Computer Science\\
    University of California, Los Angeles\\
	\texttt{shi.feng@cs.ucla.edu} \\
		\And
	Tian Han \\
    Department of Computer Science\\
    Stevens Institute of Technology\\
	\texttt{than6@stevens.edu } \\
	    \And
	Song-Chun Zhu \\
	Department of Statistics\\
	University of California, Los Angeles\\
	\texttt{sczhu@stat.ucla.edu} \\
}

\renewcommand{\shorttitle}{\textit{arXiv} Template}

\begin{document}
\maketitle

\begin{abstract}
	With the recent success of deep learning algorithms, many researchers have focused on generative models for human motion animation. However, the research community lacks a platform for training and benchmarking various algorithms, and the animation industry needs a toolkit for implementing advanced motion synthesizing techniques. To facilitate the study of deep motion synthesis methods for skeleton-based human animation and their potential applications in practical animation making, we introduce \genmotion: a library that provides unified pipelines for data loading, model training, and animation sampling with various deep learning algorithms. Besides, by combining Python coding in the animation software \genmotion\ can assist animators in creating real-time 3D character animation. Source code is available at \url{https://github.com/realvcla/GenMotion/}.
\end{abstract}


\section{Introduction}

\begin{figure}[t]
    \centering
    \includegraphics[width=\textwidth]{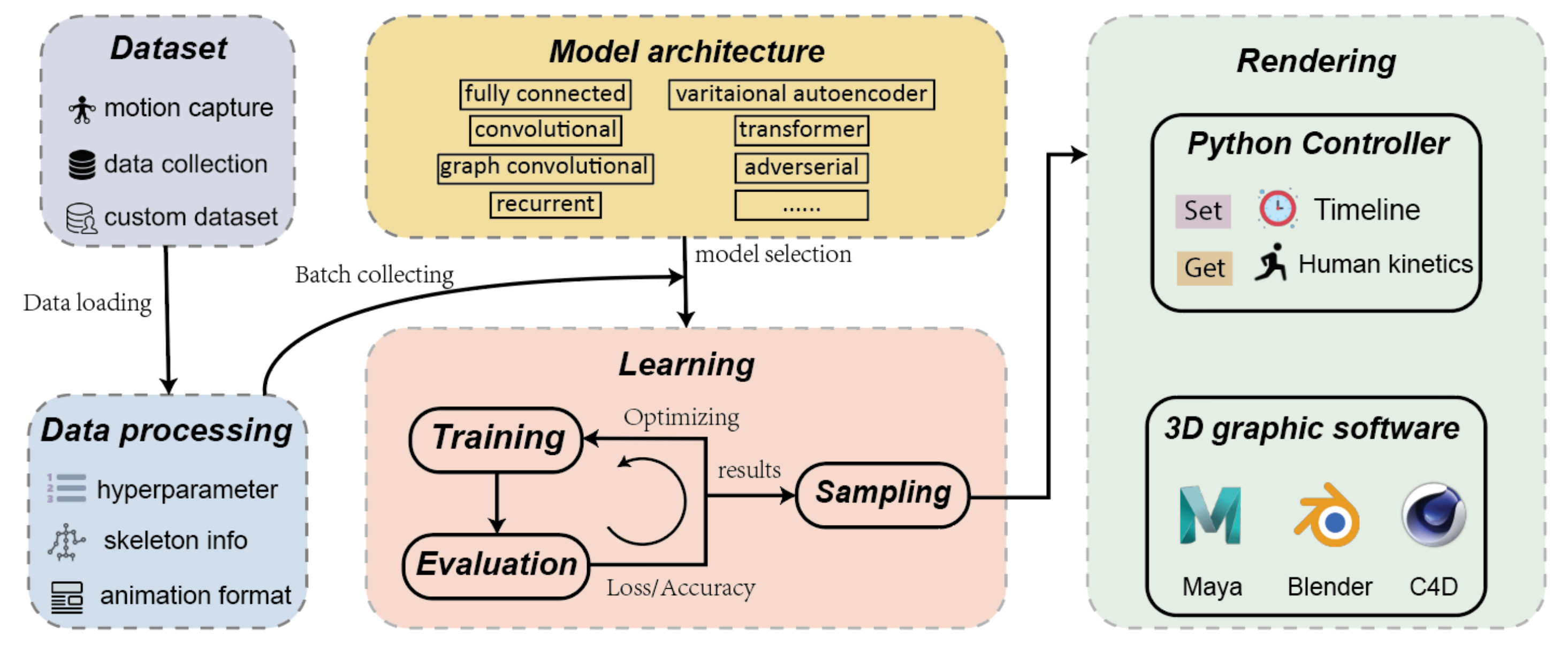}
    \caption{\textbf{library overview}. (1) \genmotion\ supports a wide range of skeleton-based motion datasets from motion captures and other collections, and it allows users to set up their custom dataset. (2) The data processing modules store the hyperparameters and skeleton information for different datasets and provide different utility functions for parsing various types of animation formats. (3)~\genmotion\ collects many deep learning models for motion generation. (4)~The learning module performs as model training and evaluation. The best model is saved for sampling animation. (4)~\genmotion\ provides the Python controller to help render real-time animation by communication with the \textit{socketserver} module in the 3D graphic software.}
    \label{fig:my_label}
\end{figure}
Computer-assisted human character animation plays an important role in making entertainment content for video games, virtual reality, and fiction films~\cite{mourot2021survey}. Over the years, the rapid evolution of data-driven methods based on machine learning has facilitated automated character animation, making it possible for animators to automatize some of the tough processes of generating human motion sequences. Traditional approaches for automated character animation include hidden Markov models~\cite{tanco2000realistic, ren2005data}, Gaussian processes~\cite{wang2007gaussian, fan2011gaussian}, and restricted Boltzmann machines~\cite{taylor2009factored}. Recently, many deep learning algorithms have been explored to help synthesize human motion animations by framing the poses or gestures~\cite{pavllo2019modeling}. Deep learning-based approaches may handle the generation of complex human motion and provide cheaper and faster animation synthesizing techniques with more fidelity and creativity~\cite{mourot2021survey}.  

However, to the best of our knowledge, due to the discrepancies in the parameterization of body rigging among different datasets and the differences in the data processing of animation clips among various algorithms, there is still a lack of a library for implementing and benchmarking advanced motion synthesizing algorithms on multiple datasets. Besides, the animation community still lacks a toolkit that allows the state-of-the-art animation generation models to be applied to time-consuming, expensive, and tedious animation projects~\cite{ribet2019survey}. 

We introduce \genmotion, an extensible, easy-to-use, and comprehensive Python library for deep learning-based human character motion generation tasks. We build \genmotion\ with an organized modular architecture that combines data processing, model training/testing, motion rendering, and real-time motion synthesizing. \genmotion\  naturally endows researchers an easy way of benchmarking their algorithms and offers animators the convenience of bringing the state-of-the-art motion synthesizing models into practice.


\section{Library overview}

\genmotion\ is developed by the Center for Vision, Cognition, Learning, and Autonomy at the University of California, Los Angeles. It enables easy data loading and experiment sharing for synthesizing skeleton-based human animation with the Python API. This section briefly describes how \genmotion\ can be used in several common scenarios of motion generation tasks. \\ 

\noindent \textbf{Working with datasets:} We integrate multiple skeleton-based human motion datasets in \genmotion. For datasets with different parameterization of the body, we include documents for meta-data descriptions and visualization tools to illustrate the characteristics of each dataset. Datasets covered include, but not limited to, motion captures~\cite{muller2007documentation, rogez2016mocap}, human activity analysis~\cite{shahroudy2016ntu}, and human motion database collection~\cite{mahmood2019amass}. Besides, to facilitate a contribution to the community, \genmotion\ also provides detailed instructions for users to upload and pre-process their custom datasets.\\

\noindent\textbf{Benchmarking the state-of-the-arts:} To encourage related research in human motion generation and retrieve empirical results from most advanced methods, \genmotion\ reproduces the training procedure of character motion generation methods by reusing the cleaning the code from official implementation~\cite{petrovich2021action, guo2020action2motion}, updating and revising the code~\cite{habibie2017recurrent}, or re-implementing the technique based on the description of the paper~\cite{liu2021motion}. One goal of \genmotion\ is to simplify the comparison of the performance between different algorithms. \genmotion\ also enables an easy setup for conducting experiments on various datasets by simply modifying settings of the hyperparameter.\\

\noindent\textbf{Rendering:} 
We provide a communication interface, i.e., client and server interaction, with the 3D modeling software in \genmotion\ to achieve real-time animation rendering and sampling. At present, 3D animation generated from deep-learning models can be rendered in several popular character animation making tools: \textsc{Autodesk Maya}, \textsc{Maxon Cinema 4D}, and \textsc{Blender}. 

\section{Use case}

Another essential goal of \genmotion\ is to apply the state-of-the-art motion synthesizing to practical animation making. Here, we list a few common use cases of our library. The full demo and tutorial for them can be found at \url{https://genmotion.readthedocs.io/en/main/genmotion_tutorials.html}\\

\begin{figure}[t]
    \centering
    \includegraphics[width=\textwidth]{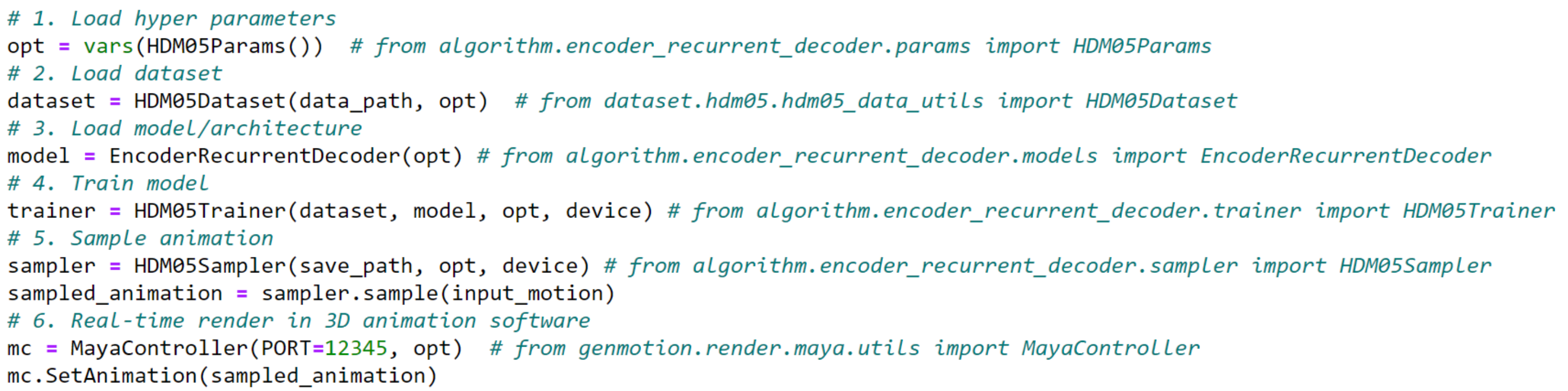}
    \caption{\textbf{Code sample}. (1) \genmotion\ stores the hyperparameters for loading a dataset. (2) \genmotion\ provides the dataloader for loading the motion dataset from $data\_path$. (3) (4) After setting up the model architecture, users can train the model easily on the $device$(CPU or GPU). (5) (6) User may can sample animation clips and render them in the animation software simultaneously.}
    \label{fig:example}
\end{figure}
\noindent \textbf{Character pose prediction:} given the input data as historical skeleton-based animation, the animation sampler in \genmotion\ helps predict geometric and motion information of human body parts, which can apply to a wide range of applications for filming industry, human-computer interaction,
motion analysis, augmented reality, and virtual reality. \\

\noindent \textbf{Conditioned Human motion synthesis}:  
Taken a semantic action label like \textit{walk}, the conditional animation sampling module in \genmotion\ generates a number of realistic 3D human motion sequences, which helps improving animation variability according to the physical environment \cite{holden2017phase} or character-scene interactions \cite{starke2019neural} \\

\noindent \textbf{Multi-character animation}: To meet the growing demands of multi-character animation such as animation in social activities \cite{shu2016learning}, the Python controller in \genmotion\ communicates with the \textit{socketserver} module in 3D animation software with a unique namespace for each character in one scene, providing an automated solution to control the motion of multiple characters. 

\section{Development and maintenance}

The project is developed by the 
 Center for Vision, Cognition, Learning, and Autonomy at University of California, Los Angeles. \genmotion\ is developed publicly through \textit{Github} with an issue tracker to report bugs and ask questions. Documentation consists 
of tutorials, examples, and API documentation.
The third-party packages include Pytorch~\cite{paszke2019pytorch} for the deep learning framework, Huggingface Transformers~\cite{wolf2019huggingface} for transformer architectures, and Jupyter notebook~\cite{randles2017using} for the demo and tutorial.

\section{Conclusion}
We presented \genmotion, an Python library to help researchers to easily develop their character animation synthesis methods and compare the results with recent deep learning algorithms. Full documentation is available at \url{https://genmotion.readthedocs.io/}.

\bibliographystyle{unsrt}
\bibliography{citations}

\end{document}